\documentclass[aps, prl, preprint, superscriptaddress, a4paper, 11pt, floatfix]{revtex4-1}
\usepackage{graphicx}
\usepackage{epstopdf}
\usepackage{amsmath}

\usepackage{silence}
\usepackage[usenames,dvipsnames]{color}
\usepackage{colortbl}
\newcommand{\TV}[1]{} 

\usepackage{float}

\usepackage[bookmarks=True,bookmarksopen=True]{hyperref} 
\hypersetup {
    unicode=false,          
    pdftoolbar=true,        
    pdfmenubar=true,        
    pdffitwindow=false,     
    pdfstartview={FitH},    
    pdftitle={Probing electronic wavefunctions by all-optical attosecond interferometry},    
    pdfauthor={Doron Azoury},     
    pdfcreator={Doron Azoury},   
    colorlinks=true,       
    linkcolor=BlueViolet,          
    citecolor=BlueViolet,        
    filecolor=Black,      
    urlcolor=Black           
}

\graphicspath{{./}}  

\usepackage{geometry}

\begin{document}

\newgeometry{left=3cm,right=3cm,top=2.5cm,bottom=2.5cm}

\title{Probing electronic wavefunctions by all-optical attosecond interferometry}

\author{Doron Azoury}
\affiliation{Department of Physics of Complex Systems, Weizmann Institute of Science, Rehovot 76100, Israel}

\author{Omer Kneller}
\affiliation{Department of Physics of Complex Systems, Weizmann Institute of Science, Rehovot 76100, Israel}

\author{Shaked Rozen}
\affiliation{Department of Physics of Complex Systems, Weizmann Institute of Science, Rehovot 76100, Israel}

\author{Barry D.~Bruner}
\affiliation{Department of Physics of Complex Systems, Weizmann Institute of Science, Rehovot 76100, Israel}

\author{Alex Clergerie}
\affiliation{Universit\'{e} de Bordeaux - CNRS - CEA, CELIA, UMR5107, F-33405 Talence, France}
\affiliation{Universit\'{e} de Bordeaux - CNRS, ISM, UMR5255, F-33405 Talence, France}
\author{Yann Mairesse}
\affiliation{Universit\'{e} de Bordeaux - CNRS - CEA, CELIA, UMR5107, F-33405 Talence, France}

\author{Baptiste Fabre}
\affiliation{Universit\'{e} de Bordeaux - CNRS - CEA, CELIA, UMR5107, F-33405 Talence, France}

\author{Bernard Pons}
\affiliation{Universit\'{e} de Bordeaux - CNRS - CEA, CELIA, UMR5107, F-33405 Talence, France}

\author{Nirit Dudovich}
\affiliation{Department of Physics of Complex Systems, Weizmann Institute of Science, Rehovot 76100, Israel}

\author{Michael Kr\"uger}
\affiliation{Department of Physics of Complex Systems, Weizmann Institute of Science, Rehovot 76100, Israel}

\date{April 20, 2018}



\begin{abstract}

{\bf

Photoelectron spectroscopy is a powerful method that provides insight into the quantum mechanical properties of a wide range of systems. The ionized electron wavefunction carries information on the structure of the bound orbital, the ionic potential as well as the photo-ionization dynamics itself. While photoelectron spectroscopy resolves the absolute amplitude of the wavefunction, retrieving the spectral phase information has been a long-standing challenge. Here, we transfer the electron phase retrieval problem into an optical one by measuring the time-reversed process of photo-ionization -- photo-recombination -- in attosecond pulse generation. We demonstrate all-optical interferometry of two independent phase-locked attosecond light sources. This measurement enables us to directly determine the phase shift associated with electron scattering in simple quantum systems such as helium and neon, over a large energy range. In addition, the strong-field nature of attosecond pulse generation resolves the dipole phase around the Cooper minimum in argon through a single scattering angle, along with phase signatures of multi-electron effects. Our study bears the prospect of probing complex orbital phases in molecular systems as well as electron correlations through resonances subject to strong laser fields.

}

\end{abstract}


\maketitle


\section*{Introduction}

\begin{figure}[!htb]
\begin{center}
  \centering{\includegraphics*[width=0.8\columnwidth]{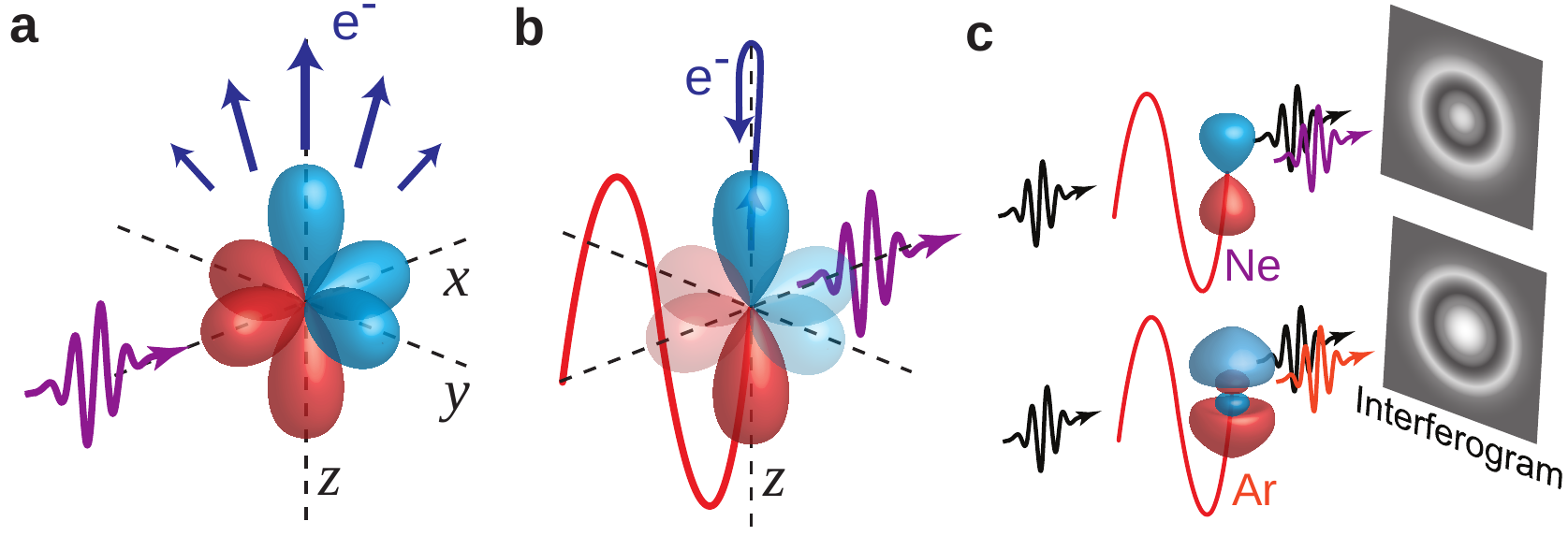}}
  \caption{{\bf Phase measurement scheme using XUV-XUV interferometry.} {\bf a}, Electrons (blue arrows) emitted from an atom after absorption of an XUV photon (violet) carry phase information on the orbital structure of the atom. Measuring the emitted electron wavepacket and its spectral phase usually averages over all initial orbitals and all emission angles. {\bf b}, High-harmonic generation provides access to the inverse process -- photo-recombination of the recolliding electron. In contrast to photo-ionization, tunneling may select specific states, according to the polarization of the fundamental field (red). Moreover, only a single recombination angle contributes, dictated by the recollision angle. {\bf c}, All-optical differential XUV-XUV interferometry scheme. High-harmonic generation of two different gases (here for example, neon (purple) and argon (orange) atoms) driven by the same infrared driving laser field interfere, one at a time, with a common phase locked reference XUV field (black) generated at an independent source. Subtraction of the interference phase provides a differential measurement of their dipole phases.}
  \label{Concept}
\end{center}
\end{figure}

Photo-ionization offers a unique insight into quantum mechanical phenomena in matter, by projecting bound wavefunctions into ionized electronic wavefunctions which can be imaged experimentally. The complex properties of the photo-electron wavefunction are dictated by all steps of the light-matter interaction -- the initial state, the properties of the ionizing radiation as well as the complex phase accumulated by the departing electron as it interacts with the Coulomb binding force. Photo-electron spectroscopy provides a direct measurement of the amplitude of the ionized wavefunction (Figure~\ref{Concept}a), but does not directly resolve its phase.

In order to recover the phase information of the ionized wavefunctions, two main strategies have been applied. In the first approach, the angular dependence of the photoelectron spectra can be used to retrieve the phase information of the ionized wavefunction, \textit{at a given electron energy}. These so-called ''complete'' photo-ionization experiments require recording photoelectron angular distributions for linear and circular ionizing radiation, and for fixed-in-space molecules using coincidence photo-electron photo-ion spectroscopy~\cite{Becker1998,motoki2002complete,marceau2017molecular}. A recent two-dimensional momentum-resolved photo-ionization study enabled the isolation of a continuum wavefunction through an atomic resonance and, in addition, resolves its phase by interference with a reference wavefunction~\cite{Villeneuve2017}. In the second approach, photoelectron interferometry methods such as the reconstruction of attosecond beating by interference of two-photon transitions (RABBITT) and streaking techniques, based on nonlinear light-matter interaction, can be used to measure the variation of the spectral phase with energy~\cite{Paul2001,MairesseScience2003,Schultze2010}. These schemes were applied to reveal phase information in photoionization of atoms~\cite{Klunder2011,Maansson2014double,Guenot2014,Sabbar2015}, molecules~\cite{Haessler2009} or solids~\cite{Cavalieri2007}. However, these approaches provide a partial measurement of the phase: they record the derivative of the phase with respect to energy -- associated with the well-known Wigner time delay~\cite{Wigner1955} -- while the absolute phase remains inaccessible.

We can transfer the photo-electron phase retrieval problem into an optical one by means of photo-recombination. Photo-recombination, the time-reversed process of photo-ionization, is an inherent part in the mechanism of high-harmonic generation (HHG)~\cite{Corkum1993}. Here, under the influence of a strong laser field an electron is liberated by tunneling ionization, propagates in the laser field and is driven back to the parent ion. Recombination of the electron with the ion leads to the emission of photons in the extreme ultraviolet (XUV) regime. While strong-field-induced photo-recombination can be considered to be the time-reversed mechanism to photo-ionization, it introduces two fundamental differences (Figure~\ref{Concept}b). First, the photo-ionization process integrates over all initial states while the tunneling process selects specific initial states, dictated by the strong field's polarization~\cite{Young2006}. Second, in photo-ionization the emitted photo-electron momentum spreads over a large angular range, while photo-recombination projects the returning electronic wavepacket in a well-defined direction, thus providing a high degree of angular selectivity~\cite{Shafir2009atomic}.

In this paper we demonstrate a differential phase measurement of the photo-recombination process by all-optical linear XUV interferometry. Our measurement exploits both the state selectivity by tunnel ionization and the angular resolution provided by the photo-recombination process, and obtain a direct insight into the fundamental quantum properties of the photoelectron wavefunctions. A complete and accurate interferometric measurement is based on two fundamental components -- an independent control over each arm's signal and the ability to manipulate their relative delay with interferometric stability. Previous realizations of optical XUV interferometry based on two phase-locked HHG sources have been introduced within a \textit{single interacting medium}, either by splitting the fundamental laser field into two focal spots in the medium~\cite{ZernePRL1997,Kovacev2005,Bellini2006} or by driving HHG in a gas mixture~\cite{BertrandNPhys2013,kanai2007destructive}. The former method is limited to a single species due to the proximity of the foci, whereas the latter provides only a single point in the interferogram. Here, we demonstrate a complete interferometric measurement, schematically described in Figure~\ref{Concept}c. First, we fully decouple the two arms of the interferometer, allowing independent control over all aspects of the interaction -- both the matter under scrutiny and the basic properties of the interacting light. Second, we measure a complete interferogram by scanning the relative delay between the two sources with attosecond precision, over a large energy range. Performing a time-resolved interferometric measurement is of prime conceptual importance: the phase is extracted in the Fourier domain, allowing its decoupling from systematic or amplitude errors while reducing the statistical phase uncertainty to values as low as 0.1\,rad.

\section*{Results}
\begin{figure}[!htb]
\begin{center}
  \centering{\includegraphics*[width=0.75\columnwidth]{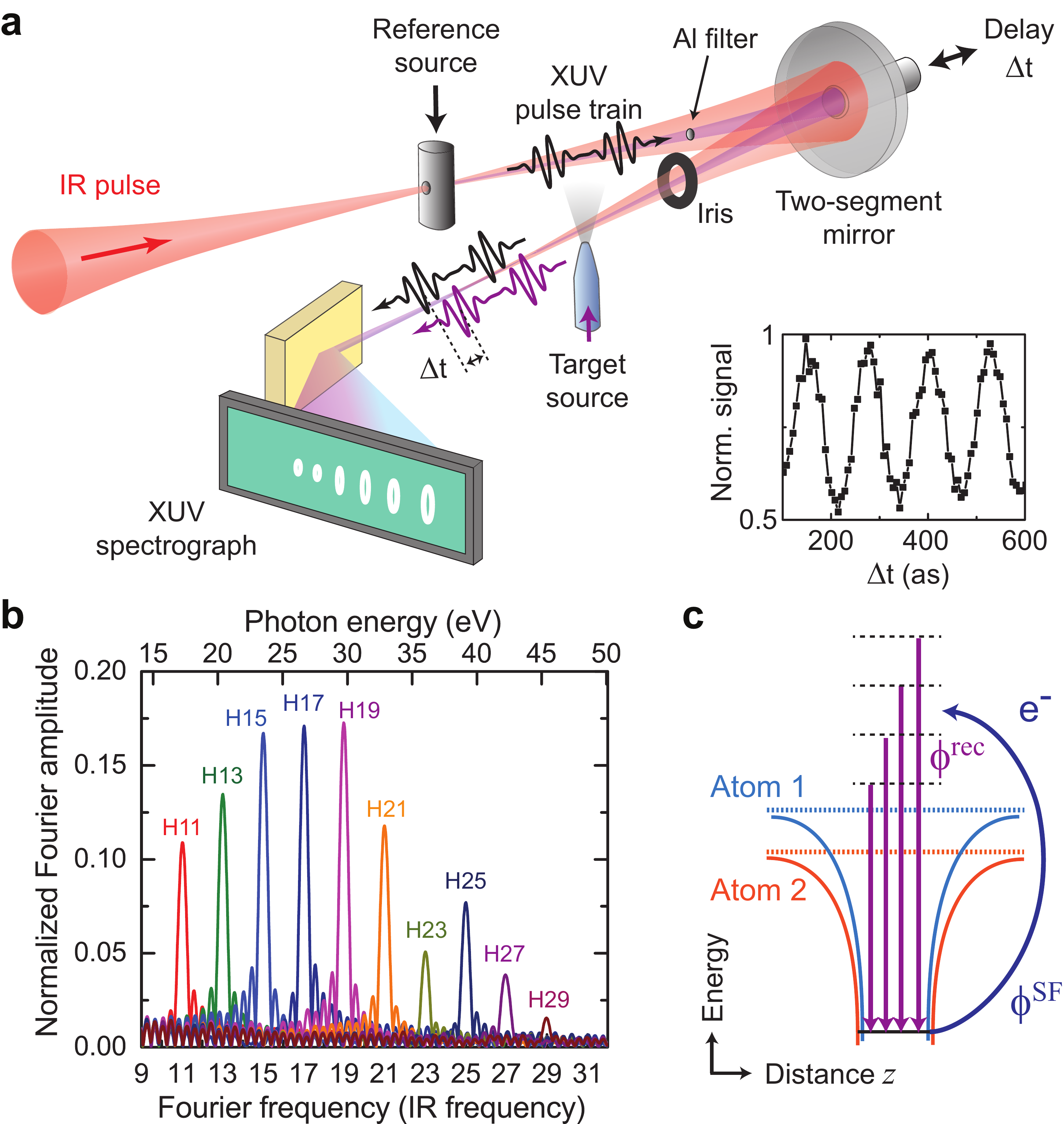}}
  \caption{{\bf XUV-XUV interferometry.} {\bf a}, Experimental setup. The IR beam is focused into the first gas cell where the reference APT is generated (black pulses). A thin Al foil transmits the APT while the inner part of the IR beam is blocked. Both beams are refocused into a second gas source by a curved two-segment mirror where the target APT is independently generated by the remaining part of the IR beam. The temporal delay $\Delta t$ between the reference and the target APTs (illustrated by the red and purple pulses, respectively) is controlled by moving the inner segment of the concave focusing mirror. An XUV spectrograph resolves the interference of the APT beams. The inset shows an example of the interference signal, measured at harmonic 21. {\bf b}, Fourier amplitudes extracted from a delay scan where argon is used both as reference and target gas. Harmonics 11 to 29 oscillate solely at their own frequency. {\bf c}, Schematic diagram of the HHG process in two different target atoms. A strong laser field leads to ionization of an electron and accelerates it back to its parent ion (blue arrow). Upon recollision, the electron undergoes photo-recombination back to the ground state while the excessive energy is emitted as high harmonics (purple arrows) of the fundamental driving field. The phase of the emitted radiation corresponds to the sum of the strong-field induced phase $\phi^\mathrm{SF}$ and the photo-recombination dipole phase $\phi^\mathrm{rec}$. The differential interferometric measurement allows us to extract the phase difference between the harmonics that correspond to dipole transitions between continuum states (dashed black lines) and the target's ground state (solid black line).}
  \label{scheme}
\end{center}
\end{figure}

Our experimental setup consists of two main stages (Figure~\ref{scheme}a, for a detailed description see Methods). In the first stage, a reference attosecond pulse train (APT) is generated by focusing an intense infrared (IR) laser pulse into a gas cell. This APT is common to all measurements, and provides a stable reference. The IR and APT beams co-propagate and are refocused into the target gas by a curved two-segment mirror. The refocused IR beam generates a second APT, phase-locked with the reference APT, which can be controlled in delay $\Delta t$ and direction using the two-segment mirror. Finally, the XUV beams from the two APTs interfere on an XUV spectrograph. When both APT beams are temporally and spatially overlapped, we observe clear interference fringes in each of the high harmonics on the spectrograph. Scanning $\Delta t$ leads to modulation of the fringe patterns over a large spectral range, directly visible in the spectrogram. A Fourier analysis (see Figure~\ref{scheme}b) of the delay-dependent intensity reveals that each harmonic oscillates according to its fundamental frequency, indicating the linear nature of the XUV-XUV interference. The phase associated with each Fourier component provides a direct measurement of the relative spectral phase between the constant reference APT and the target APT for each harmonic order.

\begin{figure}[!htb]
\begin{center}
  \centering{\includegraphics*[width=0.9\columnwidth]{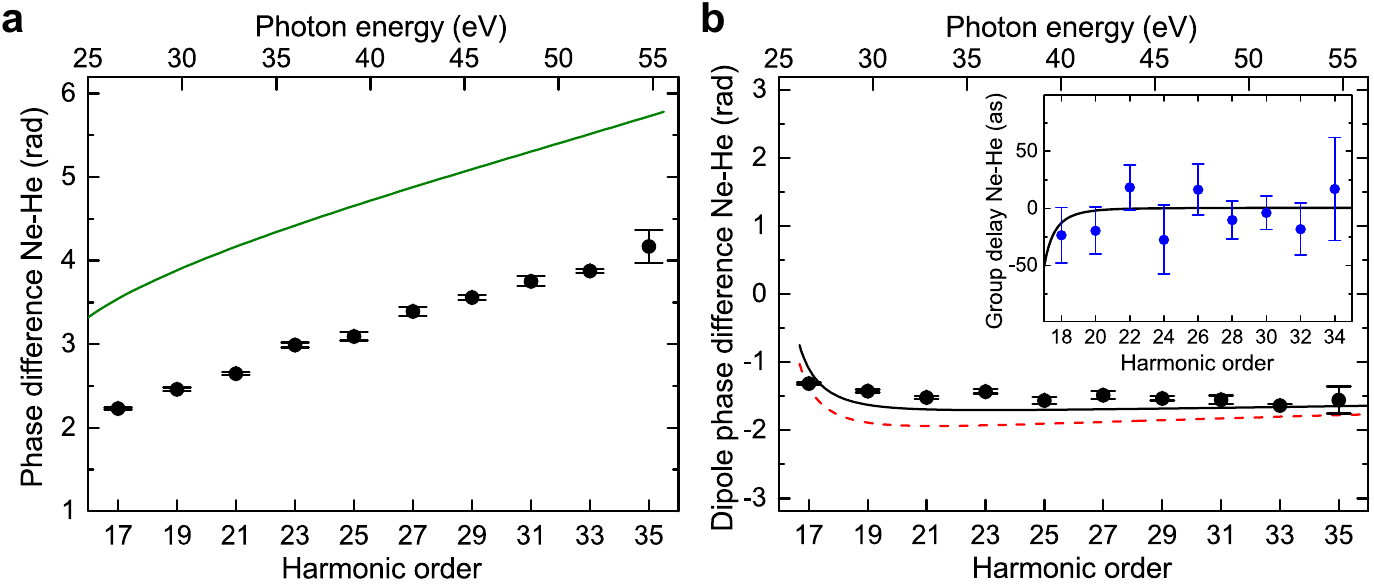}}
   \caption{{\bf Differential phase measurements of neon and helium.} {\bf a}, Phase difference of neon and helium as a function of harmonic order (black dots; error bars: standard error of the mean). The green line represents a quantum calculation of the strong-field contribution to the phase difference, $\Delta\phi^\mathrm{SF}(\Omega)$, at a laser intensity of $2.4 \pm 0.2 \times 10^{14}\,\mathrm{W\,cm}^{-2}$. The experimental data is subject to a systematic time-domain error of 5\,as due to the piezo drift correction, translating into a phase error of $\pm 0.3$\,rad for H25 (see Methods). {\bf b}, Experimental phase difference of neon and helium (as in {\bf a}) after subtracting of the strong-field phase contribution. The black line corresponds to the exact calculation of the dipole phase difference
$\Delta\phi^\mathrm{rec}_\mathrm{Ne-He}(\Omega)$, whereas the dashed red line corresponds to solely the partial wave phase shifts (see text). The proximity of the partial wave phase shifts to the exact calculation and the experimental results show that within a good approximation we can directly measure the partial wave phase shifts. The data is subject to systematic errors due to drift (see {\bf a}) and uncertainty in determining the IR intensity (H25: $\pm 0.1$\,rad). The inset shows the group delay of neon and helium (blue circles: experimental data with standard error of the mean; black solid line: theory).}
  \label{fig3}
\end{center}
\end{figure}

We perform a differential phase measurement by generating the reference APT in molecular nitrogen and interfering it with a target APT generated either in neon or helium under identical experimental conditions. In each measurement we scan the delay $\Delta t$ and extract the Fourier phase of the signal oscillations as a function of harmonic order (see Methods). The Fourier phase represents the relative spectral phase difference between the two arms of the interferometer: $\phi(\Omega)=\phi_\mathrm{tar}(\Omega) - \phi_\mathrm{ref}(\Omega)$, where $\Omega$ is the harmonic frequency. Alternating the target species while keeping the reference arm constant cancels the reference phase upon subtraction of the two interference phases. Accordingly, we are left with the absolute phase difference between neon and helium at each harmonic frequency $\Delta\phi_\mathrm{abs}(\Omega)=\phi_\mathrm{Ne}(\Omega)-\phi_\mathrm{He}(\Omega)$ (see black dots in Figure~\ref{fig3}a). This phase measurement is fundamentally different compared with photoelectron interferometry approaches, such as RABBITT or streaking. These methods record the spectral phase derivative while our scheme resolves the absolute spectral phase difference between the harmonic radiation of the two atomic species. Furthermore, in the photoelectron measurements an additional laser field is required to produce the interference pattern, such that the measurement intrinsically introduces an additional phase shift~\cite{Klunder2011,Dahlstrom2012a}. Our method provides a direct measurement of the phase through linear time-resolved XUV-XUV interferometry.

The phase of the emitted harmonics encodes all the steps of the interaction: ionization and propagation, both driven by the strong IR field, along with the photo-recombination step. Hence, as illustrated in Figure~\ref{scheme}c, we can express each harmonic phase as the sum of the strong-field-induced phase $\phi^\mathrm{SF}$ and the photo-recombination dipole phase $\phi^\mathrm{rec}$. Accordingly, the differential phase measurement provides $\Delta\phi_\mathrm{abs}(\Omega)=\Delta\phi^\mathrm{SF}(\Omega)+\Delta\phi^\mathrm{rec}(\Omega)$. The strong-field contribution $\Delta\phi^\mathrm{SF}(\Omega)$ can be captured quantitatively by the well-established strong-field approximation (SFA)~\cite{Lewenstein1994,Varju2005,Shafir2012}. The green line in Figure~\ref{fig3}a represents the phase difference $\Delta\phi^\mathrm{SF}(\Omega)$ between neon and helium according to the SFA. Clearly, the measured phase differences show a significant gap compared to the SFA phase calculation. Importantly, this gap does not affect the Wigner delay, and it is visible only due to the ability to measure the absolute phase differences. Next, we extract the quantity of interest -- the photo-recombination phase. For that, we employ the factorization of the HHG mechanism~\cite{LePRA2008,JinPRA2011,FrolovPRA2011} and subtract the SFA phase from the measured phase differences, isolating the photo-recombination phase (see black dots in Figure~\ref{fig3}b). In order to experimentally justify the factorization of the total phase difference into the strong-field and dipole parts, we performed a systematic study by measuring the phase differences of two atomic species for different IR intensities. Once we calibrate the SFA phase associated with each IR intensity, the different measurements overlap, confirming the validity of the SFA calculation and enabling us to isolate the recombination dipole phase difference (see Supplementary Material).

As discussed in the introduction, our approach provides a two-fold advantage over photo-ionization-based methods, stemming from the strong-field nature of the HHG mechanism. First, when an electron is ionized via tunneling from a group of orthogonal states (such as $2p_x$, $2p_y$ and $2p_z$ in neon) of closed-shell atoms, the state parallel to the electric field is much more efficiently ionized~\cite{Ammosov1986,Young2006}, as illustrated in Figure~\ref{Concept}b. Such initial state selectivity is generally not available in single-photon photo-ionization experiments where all the orthogonal orbitals will contribute to the dipole transition, as illustrated in Figure~\ref{Concept}a. Fortunately, the initial state selectivity translates into a final state selectivity of the photo-recombination process since in HHG, as a parametric process, the initial and final states are identical. Second, the strong-field-driven recollision process sets a strong spatial filter, allowing interaction through an extremely narrow recollision angle~\cite{Shafir2009atomic}. Therefore, we avoid averaging over the angular distribution of the recombination dipole. Based on the state and angle selectivity, we calculate the photo-recombination phase of both atomic species using a Hartree-Fock/X$\alpha$ approach (see Methods). Figure~\ref{fig3}b shows striking agreement between the experimentally extracted dipole phase differences and the calculated ones (indicated by the black line) over a large energy range between harmonics 17 to 35. 

A deeper insight into the information encoded in our measurement can be obtained by using partial wave expansion of the continuum wavefunctions in order to
describe the dipole phases of helium and neon in terms of fundamental partial-wave phase shifts $\sigma_{k,l}$, where $k$ and $l$ are the photo-electron momenta and
angular momentum quantum numbers, respectively (see Methods). Experimental access to these phase shifts on an absolute scale is of paramount importance as they
characterize the underlying ionic potential and further dictate the magnitude of scattering amplitudes \cite{Bransden}. Simple dipolar selection rules
lead in helium to a dipole phase $\phi_\mathrm{He}^\mathrm{rec}(k)$ directly linked to $\sigma_{k,1}$ according to
$\phi_\mathrm{He}^\mathrm{rec}(\Omega)=\sigma_{k,1}-\frac{\pi}{2}$.
On the other hand, the well-known (Fano) propensity rule \cite{Fano1985}, which favours $l \rightarrow l+1$ transitions, yields
$\phi_\mathrm{Ne}^\mathrm{rec}(\Omega) \simeq \sigma_{k,2}-\pi$ in neon.
We can therefore interpret our dipole phase difference as a direct measurement of the partial-wave
phase shifts, {\textit i.e.} $\Delta\phi_\mathrm{Ne-He}^\mathrm{rec}(\Omega) \simeq \sigma_{k',2}-\sigma_{k,1}-\frac{\pi}{2}$, where $k'$ and $k$ are
related by the ionization potential difference of neon and helium. The phase shift difference is illustrated in Fig.~\ref{fig3}b by the red dashed line.
Note that this is an absolute measurement of the phase, accordingly no vertical offset was artificially added between experiments and theory. The close proximity between the phase shift and exact dipole phase differences confirms our interpretation -- indeed, our measurement provides an extremely accurate and sensitive probe of the partial-wave phase shifts. In addition, the total error of measured phases (about $\pm 0.4$\,rad for harmonic 25) then sets an upper boundary for the difference in ionization dipole phase, associated with the first step of the HHG mechanism, between the two atomic species.


We have shown that our interferometer can reveal phase shifts between the wavefunctions ionized from the two lightest noble gases. In the following step we investigate the possibility of probing a more complex quantum mechanical system: the argon atom, whose photo-ionization produces a characteristic structural feature in the energy range of our measurement -- a Cooper minimum \cite{Cooper1962}. This minimum, located around 50\,eV, is induced by the shape of the atomic potential. In our study, we demonstrate the ability to directly resolve the absolute phase associated with it via XUV-XUV interferometry.

\begin{figure}[!htb]
\begin{center}
  \centering{\includegraphics*[width=0.9\columnwidth]{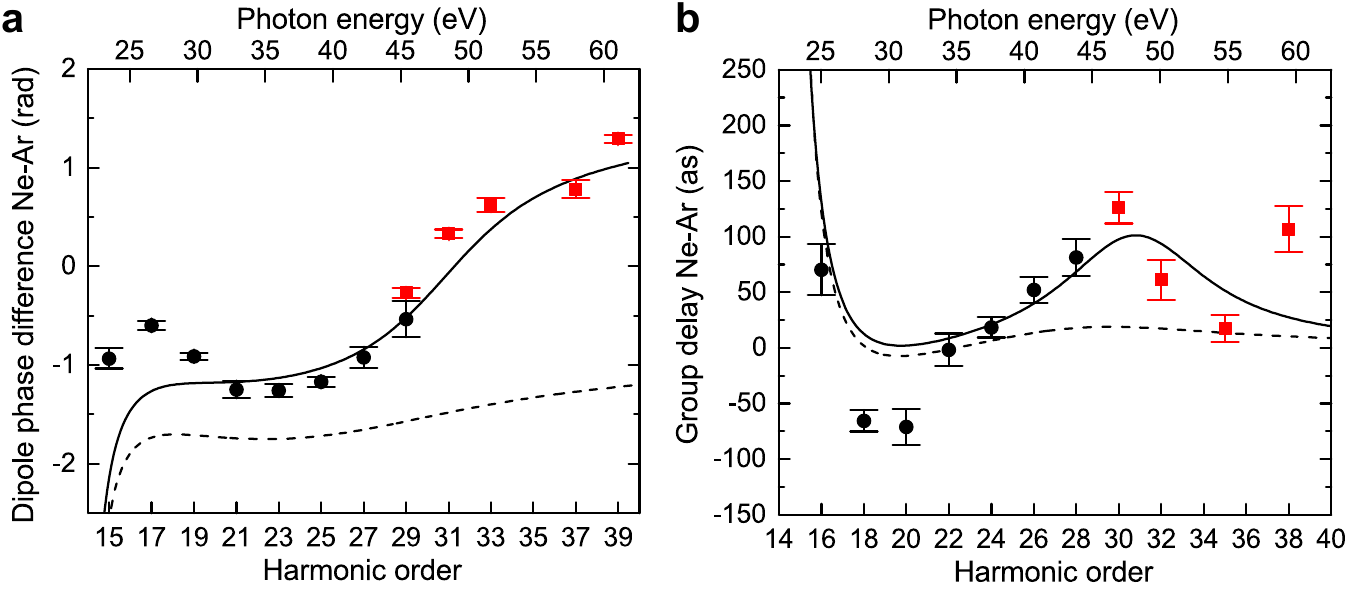}}
   \caption{{\bf Differential phase measurements of argon and neon.} {\bf a}, Measured photo-recombination dipole phase difference of neon and argon. Black dots correspond to the intensity average of 6 measurements with argon as the source for the reference APT (see Supplementary Material). The red dots display the measurements with neon as the source for the reference. The error bars show the statistical error of the mean. The solid and dashed lines represent the calculated photo-recombination dipole phase differences without and with averaging over scattering angle, respectively. The data is subject to systematic errors due to drift (see Methods) and uncertainty in determining the IR intensity (H25: $\pm 0.1$\,rad). {\bf b}, Group delay between neon and argon. Dashed and solid lines as in {\bf a}.}
  \label{fig4}
\end{center}
\end{figure}

We performed a differential measurement by alternating the target source between neon and argon. For the reference APT, we used argon or neon as a source for measurements between harmonics 15 to 29, and harmonics 23-39, respectively. We exploit the relatively flat dipole phase of neon in order to resolve the dipole phase associated with the more complex structure of the argon atom. Figure~\ref{fig4}a shows the measured phase difference between neon and argon after the subtraction of the SFA phase contribution, as in the neon and helium measurement. The black and the red symbols correspond to measurements with argon or neon as the source for the reference APT, respectively. We found full consistency in the common energy range (harmonics 23-29; see Supplementary Material), indicating the reliability of the absolute phase difference measurement. The solid line corresponds to the calculated dipole phase differences, as we described above, which shows a very good agreement with the experimental results for harmonics 21 to 39. The phase deviation at harmonics 15-19 will be discussed further below.

We now focus on the role of the atomic structure in argon that is manifested in the Cooper minimum in photo-ionization spectra. This extensively studied phenomenon is created by a $\pi$-shift of the dipole phase in one of the angular momentum ionization channels. As the time-reversed process of photo-ionization, photo-recombination is subject to the same effect, leading to a local minimum in the HHG spectrum of argon~\cite{WornerPRL2009,Higuet2011}. Here, we directly measure the spectral phase over this energy range (harmonics 27-33 in Figure~\ref{fig4}a). In particular, we use the differential scheme in order to isolate the phase difference with respect to the structureless dipole phase of neon. The dipole phase calculation (solid line in Figure~\ref{fig4}a) takes into account only one scattering angle (parallel to the strong field polarization) and accounts for both angular momentum recombination channels, $s\rightarrow p$ and $d\rightarrow p$. Comparing the calculated phase with the experimentally measured results shows a good agreement. In Figure \ref{fig4}a, we compare our measurement with an angle-integrated calculation (see Supplementary Material). Clearly such an integration does not exhibit the spectral fingerprint of the Cooper minimum. Figure~\ref{fig4}b shows the group delay associated with the measured phases, along with the group delay for the single scattering angle (solid line) and the angle-averaged (dashed line) calculations. Again, the angle and state selectivity are found to be crucial in order to probe the underlying structure of argon. Recent experiments~\cite{Schoun2014} showed that HHG together with photoelectron interferometry measurements can be used in order to measure the group delay (spectral derivative) across the energy range of the Cooper minimum without angle averaging. The results in Figure~\ref{fig4}b agree well with their measurements of the group delay. In this study, the combination of the high angular selectivity, provided by the HHG process, together with the linear nature of XUV-XUV interferometry allows us to directly measure the absolute dipole phase of the Cooper minimum in argon.

In the energy range of harmonics 15-19 the measured phase differences show clear deviation with respect to the dipole calculation. We trace back the root of this deviation to the autoionization resonances of argon ($3s3p^6np$), located approximately within a photon energy range of 26.6\,eV (harmonic 17 in our experiment) to 29.3\,eV~\cite{Beutler1935}. Our dipole phase calculation, based on a single active electron approximation, cannot capture any resonance-induced phase shifts associated with electron-electron correlation. In photo-ionization measurements, these resonances induce narrow dips in the photo-electron spectrum. The spectral phase of the argon resonance was measured in a weak field environment by RABBITT~\cite{Kotur2016}. However, recent works have revealed the influence of a strong laser field on autoionization resonances and identified both broadening as well as significant spectral modifications of the resonance line shape~\cite{Wang2010,Kaldun2016observing}. We thus infer that the broad spectral response we observe, ranging over harmonics 15-19, is due to the high intensity we apply, two orders of magnitude higher than in a previous work~\cite{Wang2010}. Accordingly, the deviation of the measured phase from the single-electron phase calculation (Figure~\ref{fig4}a) can be interpreted as the induced phase shift of the strongly dressed resonances. This result can be understood as the accumulated phase of the electron captured in the autoionizing state, until it undergoes recombination. However, a quantitative analysis of this spectral response requires an advanced calculation that would account for the multi-electron nature of the resonant process occurring in the strong field environment. Finally, note that traces of $2s2p^6np$ autoionizing resonances of neon~\cite{Langer1997} should similarly arise in our phase measurements (Figures~\ref{fig3}b and~\ref{fig4}a) at harmonics 29-31. However those resonances have a significantly lower cross section than their argon counterparts~\cite{Berrah1996}, therefore they are not visible in our phase measurements, indicated by the agreement of the neon-helium results with the single electron calculations.

\section*{Conclusions and outlook}

In this paper we have presented a novel phase measurement scheme using time-resolved interferometry of two fully independent broadband XUV sources, allowing for a differential phase measurement over distinct atomic species with temporal resolution of about 6\,as. We were able to measure the absolute phase difference of the harmonic emission of neon and helium atoms over a large energy range and isolate their dipole phase difference. In particular, our results provide a direct measurement of the species-resolved partial wave phase shift of the electronic wavefunction in the continuum. Moreover, we applied our scheme to probe the argon atom, where we demonstrated a direct measurement of its complex dipole phase. The strong-field induced spatial confinement allowed us to follow the rapid variation of the dipole phase over the energy range of the Cooper minimum, visualizing the robustness of the angular momentum propensity rule in the vicinity of an atomic structure.

Looking forward, the interferometric control over two fully independent XUV sources opens the door to new directions in both attosecond metrology and control. In molecular systems, a strong laser field can initiate multi-electron dynamics, leading to population transfer, electron rearrangement during ionization or charge migration, all of which leave a clear spectral fingerprint. A direct measurement of the spectral phase associated with such phenomena will serve as a sensitive probe of the different internal channels, and will allow following their temporal evolution. Our interferometer can also be extended straightforwardly from a measurement to a control scheme. Manipulating the interference between two independent phase-locked sources opens a new route for shaping the spectral amplitude or polarization state of an attosecond beam.

\section*{Methods}

\subsection*{Experimental setup and measurement scheme}

Figure~\ref{scheme} shows a schematic sketch of the experimental setup for collinear XUV-XUV interferometry of two independent sources. An amplified Ti:sapphire laser system operated at 1\,kHz repetition rate delivers $\sim$$23$\,fs pulses at a central wavelength of 792\,nm. Focusing the beam into a continuous flow gas cell (filled with nitrogen or argon or neon) generates the reference APT. We spatially separate the co-propagating IR and APT beams by a thin aluminum filter (200\,nm thickness). Both beams are then refocused by a curved two-segment mirror (750\,mm focal length) into the target gas (continuous flow glass nozzle, orifice of approximately 10\,$\mu$m) in order to produce the target APT which interferes with the reference APT. The position of the target source with respect to the IR focus is adjusted to produce short trajectories of HHG. The inner part of the focusing mirror reflects the APT in the spectral range of 17-51\,eV. A piezo stage controls the temporal delay $\Delta t$ between the reference APT and target APT with a step size of 6.7\,as and an accuracy of about 1\,as. The IR intensity at the target gas can be adjusted independently by means of a motorized iris. The co-propagating APT beams are spectrally resolved by a flat-field aberration-corrected concave grating and recorded by a micro-channel plate detector, imaged by a CCD camera. In each scan, we varied $\Delta t$ over a range of 6.7\,fs (about 2.5 IR cycles) and recorded the spectrum at each step. We applied the differential scheme by repeating the delay scan multiple times, alternating the target source every two scans. By carefully monitoring the individual gas pressure of each target source while alternating the atomic species we ensured the repeatability of the interaction conditions in the target source. For each scan, we extracted the Fourier phase $\phi(\Omega)$ of every harmonic by averaging over the phases of all individual pixels. We only took pixels into account that exhibit a significant signal-to-noise ratio and are located in regions where the short-trajectory HHG signal dominates. We then calculate the experimental phase differences $\Delta\phi_\mathrm{abs}(\Omega)=\phi_\mathrm{Ne}(\Omega)-\phi_\mathrm{He}(\Omega)$ of neon and helium or $\Delta\phi_\mathrm{abs}(\Omega)=\phi_\mathrm{Ne}(\Omega)-\phi_\mathrm{Ar}(\Omega)$ of neon and argon, and the corresponding standard error of the mean from averaging over the phase differences of all the scan measurements. In order to estimate the slow thermal drift of the piezo (up to 50\,as per hour), we determined the temporal drift between pairs of identical measurements of every target source. We corrected for the drift, resulting in a systematic error in the experimental group delay (or equivalently, a linear phase shift). Typically, this error amounts to about 5\,as, translating into a frequency-dependent error in the phase differences of $N \cdot 0.012$\,rad ($N$: harmonic number). Both in the neon-helium and neon-argon differential measurements, we measured the absorption of the reference APT in each target gas in order to evaluate the neutral gas dispersion difference. Using tabulated values of gas dispersion~\cite{NISTfftable}, we set an upper limit of 0.05\,rad and 0.1\,rad, for neon-helium and neon-argon, respectively, which is below our experimental error. In addition we repeated the neon-argon differential measurement for different pressure levels of argon in the target source (all other parameters remained constant), and found full overlap of the phase differences over all the measured harmonics (see Supplementary Material). Therefore we conclude that in our measurements electron plasma dispersion and neutral gas dispersion induced phase distortion can be neglected.

\subsection*{Strong-field theory model}

The strong-field theory model is based on the stationary phase approximation applied to the strong-field approximation (SFA)~\cite{Lewenstein1994,Varju2005}. The strong-field phase $\phi^\mathrm{SF}$ of a harmonic with frequency $\Omega$ can be expressed in atomic units as
\begin{equation}
\phi^\mathrm{SF}(\Omega) = \operatorname{Re} \left\{ {\Omega t_1(\Omega) - \int_{t_0(\Omega)}^{t_1(\Omega)} \left( \frac{[p(\Omega) - A(t)]^2}{2} + I_\mathrm{p} \right) dt} \right\},
\label{eq:phiSF}
\end{equation}
where $t_0(\Omega)$ and $t_1(\Omega)$ are the (complex) ionization and recollision times, respectively, and $p(\Omega)$ the canonical momentum of the outgoing electron. $A(t)$ is the time-dependent vector potential of the infrared laser field (here assumed to be a continuous wave) and $I_\mathrm{p}$ the ionization energy of the atom. The integral contains both the classical action of the electron trajectory and the phase evolution of the ground state. The trajectory is defined by the parameters $t_0$, $t_1$ and $p$, which are calculated numerically for the short trajectory branch.

\subsection*{Hartree-Fock/X$\alpha$ calculation of the dipole phase}

Atomic units are used throughout this section unless otherwise stated. We assume that the whole HHG process is dictated by the dynamics of only one outer valence electron, with all other electrons remaining frozen throughout the interaction. The photo-recombination dipole in the strong-field driven $z$-direction is then
\begin{equation}
d_\text{X}^\mathrm{rec}(\Omega)=\langle\phi_0 | z | \Psi_{\bf k}^{+} \rangle
\label{dipole_X}
\end{equation}
where $\phi_0({\bf r})$ is the outer valence bound orbital of atom X and $\Psi_{\bf k}^+({\bf r})$ is the outgoing scattering state
associated with the electron returning to the core with energy $E=k^2/2$ to yield the harmonic photon of energy $\Omega=E+I_\mathrm{p}$ through recombination,
$I_\mathrm{p}$ being the ionization potential of atom X. $\phi_0$ is obtained by means of Hartree-Fock (HF) calculations, using the quantum chemistry package
GAMESS-US \cite{GAMESS} with a large-scale triple-$\zeta$ aug-cc-pVTZ underlying gaussian basis \cite{pVTZ}. $\Psi_{\bf k}^+$, normalized on the ${\bf k}$-scale, is expanded onto spherical partial waves $\psi_{klm}({\bf r})$ with definite ($l,m$)-symmetry as
\begin{equation}
\Psi_{\bf k}^+({\bf r})=\frac{1}{k}\sum_{l=0}^{\infty} \sum_{m=-l}^{l} \text{i}^l \text{e}^{\text{i}\sigma_{kl}} \psi_{klm}({\bf r}) \text{Y}^{m*}_l({\bf {\hat k}})
\label{expansion}
\end{equation}
where $\text{Y}^{m}_l$ are usual spherical harmonics and $\sigma_{kl}$ is the phase shift for electron wavevector $k$ and angular momentum $l$.
The radial part $R_{kl}(r)$ of the continuum states $\psi_{klm}({\bf r})=R_{kl}(r)\text{Y}^{m}_l({\bf {\hat r}})$ is obtained by solving numerically the Schr\"odinger equation $[\frac{1}{r^2}\frac{\partial}{\partial r}(r^2\frac{\partial}{\partial r}) - \frac{l(l+1)}{r^2}+2(E-V(r))] R_{kl}(r)=0$, where the potential $V$ is split as $V(r)=V_{e-n}(r)+V^\mathrm{(d)}_{e-e}(r)+V^\mathrm{(ex)}_{e-e}(r)$. $V_{e-n}$ is the electron-nucleus potential and $V^\mathrm{(d)}_{e-e}$ and $V^\mathrm{(ex)}_{e-e}$ are the direct (Hartree) and exchange parts of the electron-electron interaction, respectively. While $V^\mathrm{(d)}_{e-e}(r) \equiv V^\mathrm{(d)}_{e-e}({\bf r})=\int{d{\bf r'}\frac{n({\bf r'})}{|{\bf r}-{\bf r'}|}}$ where $n({\bf r'})$ is the total electron density of the multi-electron atom in its ground state, we employ for $V^\mathrm{(ex)}_{e-e}$ the so-called X$\alpha$ statistical form $V^\mathrm{(ex)}_{e-e}(r) \equiv V^\mathrm{(ex)}_{e-e}({\bf r})=-\frac{3}{2}\alpha \left( \frac{3 n({\bf r})}{\pi} \right)^{1/3}$ \cite{Slater1972}. Optimized values for the parameter $\alpha$ are tabulated in \cite{Schwarz1972}. The total potential $V(r)$ defined in this way does not present
the expected $-1/r$ asymptotic behavior; therefore we switch from the computed $V(r)$ to the so-called Latter tail $-1/r$ from $r_0$, such that $V(r_0)=-1/r_0$,
onwards \cite{Latter1955}. The $\sigma_{kl}$ phase shift encodes both the Coulombian asymptotic behavior and the short-range distortion according to $\sigma_{kl}=\sigma^\mathrm{C}_{kl}+\sigma^\mathrm{S}_{kl}$. $\sigma^\mathrm{C}_{kl}=\text{arg}\,\Gamma (l+1-\text{i}/k)$, where $\Gamma$ is the Gamma function \cite{Abramowitz}, and $\sigma^\mathrm{S}_{kl}$ is determined by matching the computed radial wavefunction $R_{kl}(r)$ to the expected asymptotic behavior $\sqrt{2/\pi}\sin (kr-l\pi/2+\sigma^\mathrm{C}_{kl}+\sigma^\mathrm{S}_{kl})/r$ at $r \rightarrow \infty$.

Once the bound and continuum wavefunctions are computed, the insertion of the partial-wave expansion (\ref{expansion}) into the expression (\ref{dipole_X})
of the dipole leads to
\begin{equation}
d_{\text{He}}^\mathrm{rec}(\Omega)=\frac{-\text{i}}{\sqrt{12 \pi}k}\text{e}^{\text{i}\sigma_{k1}}{\cal R}_{k1}
\end{equation}
for X$\equiv$He, taking into account that $\phi_0({\bf r})=R_0(r)\text{Y}^{0}_0({\bf {\hat r}})$ and ${\bf {\hat k}} // z$. ${\cal R}_{k1}$ is the radial integral
${\cal R}_{k1}=\int_0^\infty{R_0(r) R_{k1}(r) r^3 dr}$. The phase of the dipole $\phi_{\text{He}}^\mathrm{rec}(\Omega)$
then reduces to the fundamental partial-wave phase shift $\sigma_{k1}$ according to
\begin{equation}
\phi_{\text{He}}^\mathrm{rec}(\Omega)=\sigma_{k1} - \pi/2.
\end{equation}
For neon and argon, for which $\phi_0({\bf r})=R_0(r)\text{Y}^{1}_0({\bf {\hat r}})$, one obtains in a similar way
\begin{equation}
d_{\text{Ne,Ar}}^\mathrm{rec}(\Omega)=\frac{1}{\sqrt{12 \pi}k}(\text{e}^{\text{i}\sigma_{k0}}{\cal R}_{k0}-2\text{e}^{\text{i}\sigma_{k2}}{\cal R}_{k2}).
\end{equation}
The exact phase of these dipoles is
\begin{equation}
\phi_{\text{Ne,Ar}}^\mathrm{rec}(\Omega)=\text{tan}^{-1}\left[\frac{{\cal R}_{k0}\text{sin}\sigma_{k0}-2{\cal R}_{k2}\text{sin}\sigma_{k2}}
{{\cal R}_{k0}\text{cos}\sigma_{k0}-2{\cal R}_{k2}\text{cos}\sigma_{k2}}\right].
\end{equation}
Dipolar interaction generally favors $l \rightarrow l+1$ transitions, which is known as the (Fano) propensity rule \cite{Fano1985}. This translates
into our computations as ${\cal R}_{k2} \gg {\cal R}_{k0}$, but over distinct energy ranges for neon and argon. The propensity shows up over
the whole energy range considered in the experiment for neon, so that $d_\text{Ne}^\mathrm{rec}(\Omega)$ can safely be approximated by
\begin{equation}
\phi_{\text{Ne}}^\mathrm{rec}(\Omega)\simeq\sigma_{k2}-\pi,
\end{equation}
taking into account that ${\cal R}_{k2} > 0$. The propensity does not hold in argon for harmonic orders greater than $\sim 21$ because of the occurrence
of the Cooper minimum around $E=50$ eV. This impedes the isolation of $\sigma_{k2}$ of argon in the measurement of the dipole phase $\phi_{\text{Ar}}^\mathrm{rec}(\Omega)$ out from the energy range where ($3s3p^6np$) autoionizing resonances come into play.


\subsection*{Data availability}

The data of this study are available from the corresponding author on reasonable request.

\subsection*{Acknowledgements}

We thank Serguei Patchkovskii, Christian Ott and Anne Harth for helpful discussions. N.~D.~is the incumbent of the Robin Chemers Neustein Professorial Chair. N.~D.~acknowledges the Minerva Foundation, the Israeli Science Foundation, the Crown Center of Photonics and the European Research Council for financial support. M.~K.~acknowledges financial support by the Minerva Foundation and the Koshland Foundation. B.~P.~, A.~C.~, B.~F.~and Y.~M.~acknowledge the financial support from the French National Research Agency through grant ANR-14-CE32-0014 MISFITS.

\subsection*{Contributions}
N.~D.~and M.~K.~supervised the study. D.~A.~and M.~K.~designed and built the experimental setup. D.~A.~, M.~K. and O.~K.~carried out the measurements and analyzed the data. B.~P.~, A.~C.~and B.~F.~conceived and performed the theoretical calculations. D.~A.~, M.~K.~, N.~D.~, B.~P.~, B.~F.~and Y.~M.~interpreted the experimental and theoretical results. B.~D.~B.~supported the operation of the laser system. All authors discussed the results and contributed to the final manuscript.

\subsection*{Corresponding authors}
Correspondence to Nirit Dudovich (nirit.dudovich@weizmann.ac.il).

\end{document}